\documentstyle[floats,aps]{revtex}
\begin{document}
\input epsf

\draft

  \font\twelvemib=cmmib10 scaled 1200
  \font\elevenmib=cmmib10 scaled 1095
  \font\tenmib=cmmib10
  \font\eightmib=cmmib10 scaled 800
  \font\sixmib=cmmib10 scaled 667
  \skewchar\elevenmib='177
  \newfam\mibfam
  \def\mib{\fam\mibfam\tenmib}
  \textfont\mibfam=\tenmib
  \scriptfont\mibfam=\eightmib
  \scriptscriptfont\mibfam=\sixmib
\def\frac#1#2{{\textstyle{#1 \over #2}}}
\def\pz{\partial}
\def\xhi{{\raise.35ex\hbox{$\chi$}}}
\def\nd{^{\vphantom{\dagger}}}
\def\yd{^\dagger}
\def\undertext#1{$\underline{\hbox{#1}}$}
\def\Tr{\mathop{\rm Tr}}
\def\ket#1{{\,|\,#1\,\rangle\,}}
\def\bra#1{{\,\langle\,#1\,|\,}}
\def\braket#1#2{{\,\langle\,#1\,|\,#2\,\rangle\,}}
\def\expect#1#2#3{{\,\langle\,#1\,|\,#2\,|\,#3\,\rangle\,}}
\gdef\journal#1, #2, #3, 1#4#5#6{               
    {\sl #1~}{\bf #2}, #3 (1#4#5#6)}            
\def\prb{\journal Phys. Rev. B, }
\def\prl{\journal Phys. Rev. Lett., }
\def\wt{\widetilde}
\def\psit{\widetilde\psi}
\def\ie{{\it i.e.\/}}
\def\etal{{\it et al.\/}}
\def\eg{{\it e.g.\/}}
\def\vth{\vartheta}
\def\half{\frac{1}{2}}
\def\fourth{\frac{1}{4}}
\def\cH{{\cal H}}
\def\cI{{\cal I}}
\def\cO{{\cal O}}
\def\Mt{\wt T}
\def\ctn{{\rm ctn}\,}
\def\rmU{{\rm U}}
\def\rme{{\rm e}}
\def\bfr{{\bf r}}
\def\bfB{{\bf B}}
\def\zhat{{\hat{\bf z}}}
\def\EF{E\nd_{\rm F}}
\def\Ec{E\nd_{\rm c}}
\def\DH{D\nd_{\rm H}}
\def\PL{P\nd_L(T)}

\twocolumn[\hsize\textwidth\columnwidth\hsize\csname
@twocolumnfalse\endcsname
\title{Real Space Renormalization of the Chalker-Coddington Model}
\author{Daniel P. Arovas$^1$, Martin Janssen$^2$, and Boris Shapiro$^3$}
\address{
${}^1$ Department of Physics, University of California at San Diego,
La Jolla CA 92093\\
${}^2$ Institut f{\"u}r Theoretische Physik, Universit{\"a}t zu K{\"oln},
Z{\"u}lpicherstra{\ss}e 77, 50937 K{\"o}ln, Germany\\
${}^3$ Physics Department, Technion -- Israel Institute of Technology,
32000 Haifa, Israel
}

\date{\today}

\maketitle

\begin{abstract}
We study a number of hierarchical network models related to the
Chalker-Coddington model of quantum percolation.  Our aim is to describe
the physics of the quantum Hall transition.  The hierarchical network
models are constructed by combining series and parallel composition of
quantum resistors.  The localization-delocalization transition
occurring in these models is treated by real space renormalization
techniques.  Essentially, the localization-delocalization transition
is due to a competition between two one-dimensional localization
mechanisms.
\end{abstract}

\pacs{PACS numbers: 73.40.Hm, 64.60.Ak}
\vskip2pc]

\narrowtext

\section{Introduction}\label{secint}

Despite the explosion of interest in and seemingly inexhaustible richness
of the quantum Hall effect (QHE), surprisingly little progress has been
made on what is arguably the central phenomenon associated with all
incompressible quantum Hall liquids -- the quantum Hall transition itself.
By now the basic phenomenology is rather well-known: noninteracting
electrons confined to a plane, when subjected to a fixed uniform magnetic
field and a random scalar potential, exhibit quantum critical behavior at
a sequence of energy eigenvalues, where the correlation length behaves as
$\xi(E)\propto |E-\Ec|^{-\nu}$ with $\nu\approx\frac{7}{3}$ \cite{foot1,books}.
Early theories of localization in the presence of a magnetic field held
that no extended states exist in two dimensions, a result derived from
the corresponding ``unitary class'' nonlinear sigma model description
of the long-wavelength physics.  Experiments, however, unambiguously
demonstrated the existence of extended states in the quantum Hall system.
The work of Levine, Libby, and Pruisken \cite{Lev83} showed how a novel
topological term present in the sigma model could produce the
necessary critical behavior, although technical difficulties rendered the
improved sigma model ineffective in providing a quantitative description of
the transition (\eg\ critical exponents, scaling functions, {\it etc.})
Nonetheless, the language of critical phenomena provided a very useful
framework within which one could interpret experiments \cite{Pru88,Son96}.
Consider, for instance, the behavior of the Hall conductivity $\sigma_{xy}$
as a function of the variables $B$ (magnetic field), $L$ (linear system
dimension), and $\Theta$ (temperature).  Within the scaling regime, one can
write
\begin{equation}
\sigma_{xy}(B,L,\Theta)={e^2\over h} f(|B-B^*|\,L^{1/\nu},|B-B^*|
\, \Theta^{-1/z\nu})
\end{equation}
where $B^*$ is a critical magnetic field, $z$ is the dynamic critical
exponent, and $f(x,y)$ is a universal scaling function.  From this
expression, one finds the maximum slope $(\pz\sigma_{xy}/\pz B)\nd_{\rm max}$
is proportional to $\Theta^{-1/z\nu}$; the power law divergence with
decreasing $\Theta$ is in agreement with experiments, which independently
show $z=1$ \cite{expt}.  Numerical calculations \cite{Huc95} confirm the
result $\nu=2.35\pm 0.02$, in tantalizing agreement with a compelling but
alas nonrigorous argument that $\nu$ should be $\frac{7}{3}$ \cite{Mil88}.

\begin{figure} [h]
\centering
\leavevmode
\epsfxsize=7cm
\epsfysize=7cm
\epsfbox[0 0 369 370] {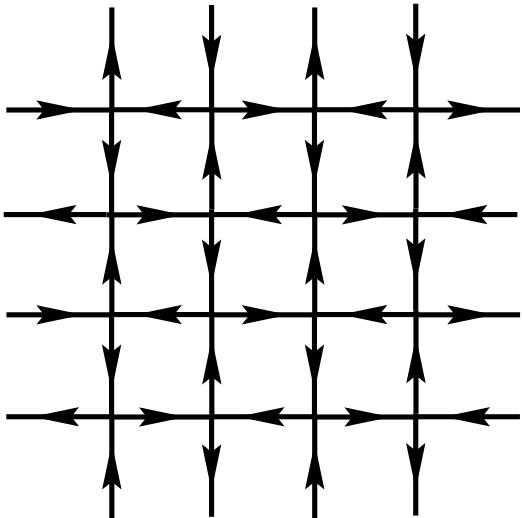}
\caption[]
{\label{CC-full} The Chalker-Coddington network model of quantum percolation.}
\end{figure}

One of the most significant developments in modeling the QHE transition
has been the advent of the Chalker-Coddington network model of quantum
percolation \cite{Cha88}.  
The relevance of {\it classical} percolation to the quantum Hall problem was
emphasized by Trugman \cite{Tru83}.  Classical electrons in a strong magnetic
field $\bfB=B\zhat$ obey the guiding center drift equations of motion
\begin{equation}
{d\bfr\over dt}=-{c\over eB}\, \zhat\times\nabla V(\bfr)
\end{equation}
where $V(\bfr)$ is the external potential.  A vivid picture emerges
in which the Fermi `sea' is analogous to a real sea covering a rough surface
in energy space whose height is described by the function $V(\bfr)$.
The corresponding quantum eigenstates are localized along equipotentials,
accruing an Aharonov-Bohm phase $\frac{e}{\hbar c}\int {\bf A}\cdot d\bfr$
which for a complete orbit is proportional to the magnetic flux encircled
(single valuedness then leads to Bohr-Sommerfeld quantization rules).
Electrons at the Fermi level either circulate clockwise (as viewed from
`above') around isolated lakes, when $\EF < \langle V\rangle$, or
counterclockwise around isolated islands, when $\EF > \langle V\rangle$.
Such states clearly are localized.  Only when $\EF=\Ec\equiv \langle V\rangle$
are states at the Fermi level extended.  As $E\to \Ec$, the equipotentials
become more and more rarefied, and their circumference increases in size
as $|E-\Ec|^{-4/3}$, an exact result.  However, the quantum eigenfunctions
are not infinitely narrow.  Rather, they have a width on the order of
the magnetic length $\ell=\sqrt{\hbar c/eB}$, and thus quantum tunneling
will occur in the vicinity of saddle points of $V(\bfr)$ \cite{Mil88,Fer87}.
It is precisely this physics which is captured by the network model.
Thus, rather than directly computing the eigenfunctions of
\begin{equation}
{\cal H}={1\over 2m}({\bf P} +{e\over c}{\bf A})^2+V(\bf r)\ ,
\label{Ham}
\end{equation}
many of which are tightly localized on peaks and valleys and hence
irrelevant to the physics of the QHE transition,
the network model ``cuts to the chase'' and simulates a network of saddle
points (see fig. \ref{CC-full}), each described by an $S$-matrix
\begin{equation}
\pmatrix{\cO\cr\cO'\cr}=S\pmatrix{\cI\cr\cI'\cr}\ ,
\ S=\pmatrix{t & r'\cr r & t'\cr}
\label{Smat}
\end{equation}
relating incoming to outgoing flux amplitudes \cite{Cha88}.  In the frame of
an incoming electron, scattering is either to the left, with probability
$T=|t|^2=|t'|^2$, or to the right, with probability $R=|r|^2=|r'|^2=(1-T)$
-- there is neither `forward' nor `backward' scattering.

When $T<\half$, scattering is predominantly to the right, corresponding to
the aforementioned clockwise motion around isolated lakes.  Hence there is
a correspondence between the energy eigenvalue $E$ for the Hamiltonian
of eq. (\ref{Ham}) and the transmission probability $T$; the quantum critical
point should lie at $T=\half$.  Randomness enters the network model
principally through phases acquired by the flux amplitudes in the course of
their propagation along each link.  These phases reflect the Aharonov-Bohm
phases accrued due to motion along equipotential segments of irregularly
varying length.  The phases on the links are therefore modeled as random
variables uniformly distributed between $0$ and $2\pi$.  Several numerical
investigations \cite{Huc95,Cha88,Lee93,Kle95} have convincingly demonstrated
the applicability of the network model to the quantum Hall transition.

In this paper, we adopt a real space renormalization group (RG) scheme which
will allow us to compute exponents and scaling functions associated with
the quantum Hall transition.  Inasmuch as real space renormalization is fraught
with uncontrolled approximations, our results will be of dubious
quantitative value.  However, we find that a simple and appealing qualitative
picture emerges from this approach, and we feel its pedagogical value alone
merits publication.  We also wish to draw the reader's attention to the recent
work of Galstyan and Raikh \cite{Gal97}, who independently developed a real
space RG approach to the network model.  Their results, of which we became
aware when this paper was in its final stages of preparation, are largely
complementary to those presented here.

\section{The Distribution $P_L(T)$}\label{dist}

\begin{figure} [h]
\centering
\leavevmode
\epsfxsize=7cm
\epsfysize=7cm
\epsfbox[18 144 592 718] {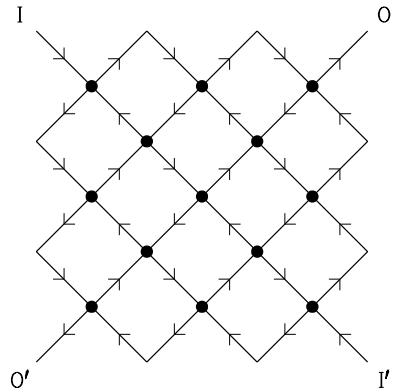}
\caption[]
{\label{net3x3} Transmission and reflection through a
$3\sqrt{2}\times 3\sqrt{2}$ network ($L=3$).  The 13 vertices are denoted
by black dots.}
\end{figure}

We commented above how a Chalker-Coddington network composed of identical
scatterers should exhibit a quantum critical point when $T=R=\half$.
Consider now a network in which the individual scattering probabilities $T$
(as well as the phases on the links) are random variables, chosen from a
distribution $P(T)$.  We shall be interested in how the distribution $\PL$
behaves with increasing cell size $L$ in the limit $L\to \infty$.
To define the transmission coefficient $T$ of a finite network, we cut out
a $\sqrt{2} L \times \sqrt{2} L$ section from an infinite network
(the lattice spacing is $a\equiv 1$), and retain only the central
$V(L)=L^2+(L-1)^2$ scattering vertices.  Fig. \ref{net3x3} shows the $L=3$
case.  Note that there are $M=2L-1$ scattering nodes across each of the main
diagonals.  Integrating over the random phases on the links, one obtains the
distribution $\PL$ for transmission to the left through the supercell.
In the limit $L\to\infty$, we expect two stable distributions, given by
$P\nd_\infty(T)=\delta(T)$ and $P\nd_\infty(T)=\delta(1-T)$.  These correspond
to bulk localized phases with $T=0$ and $T=1$, respectively.  The quantum
critical point will be characterized by an {\it unstable} distribution
$P\nd_\infty(T)=P^*(T)$.  The terms ``stable'' and ``unstable'' refer to
renormalization group flows \cite{Car96}: we shall develop an approximation
scheme by which the distribution $P\nd_{bL}(T)$ for a larger system may be
determined in terms of $\PL$. This functional relation may be represented
in terms of a set of parameters $\{X_i(L)\}$ which characterize the
distribution $\PL$ (\eg\ the coefficients in a Legendre or Chebyshev
polynomial expansion of $\PL$ in the variable $x=2T-1$):
\begin{equation}
X_i(bL)=F_i(\{X_j(L)\};b)\ .
\end{equation}
When $X_i^*=F_i(\{X_j^*\};b)$ the distribution is at a fixed point.
The eigenvalues $\{\lambda_k\}$ of the matrix
$R_{ij}=(\pz F_i/\pz X_j)\big|_{{\vec X}^*}$ determine the
relevance of the corresponding eigenvectors (scaling variables) and hence
the stability of the fixed point -- positive eigenvalues correspond to
relevant variables and an unstable fixed point, negative eigenvalues to
irrelevant variables.  The positive eigenvalues define a set of critical
exponents, $y^{-1}_k=\ln b/\ln\lambda_k$; the localization length exponent
$\nu=\ln b/\ln\lambda_{\rm max}$ is the largest of these.  One can also
define the beta function
\begin{equation}
\beta_i(\{X_j\})={d X_i\over d\ln L}={\pz F_i\over\pz b}\Big|_{b=1}\ ;
\end{equation}
the beta function vanishes at all fixed points.

From these considerations it becomes clear that one possible way to
compute the critical exponent $\nu$ would be to determine the
RG flow of $\PL$.  An exact procedure would require
the calculation of this distribution for finite networks of arbitrary size.
Computational limitations would render such attempts futile beyond even
modest values of $L$.  Instead, we shall resort to two uncontrolled
and closely-related approximations: (i) Migdal-Kadanoff (MK) bond shifting,
and (ii) replacing the square lattice with an unphysical, hierarchically
constructed lattice.  These approaches will allow us to perform the
renormalization recursively, using only simple numerical computations.

\section{Series and Parallel Propagation}\label{serpar}

The group $\rmU(2)$ of unitary $2\times 2$ matrices
may be parameterized by four angles $(\alpha,\beta,\gamma,\theta)$:
\begin{displaymath}
S= e^{i\alpha}\pmatrix{e^{i\beta}&0\cr 0&e^{-i\beta}\cr}
\pmatrix{\cos\theta &\sin\theta\cr-\sin\theta&\cos\theta\cr}
\pmatrix{e^{i\gamma}&0\cr 0&e^{-i\gamma}\cr}.
\end{displaymath}
The phase angles $\alpha$, $\beta$, and $\gamma$ can be absorbed by
the random link phases, and hence without loss of generality one may
restrict attention to $S$-matrices of the form
\begin{equation}
S=\pmatrix{\cos\theta &\sin\theta\cr-\sin\theta&\cos\theta\cr}\ .
\end{equation}
\begin{figure} [t]
\centering
\leavevmode
\epsfxsize=9cm
\epsfysize=9cm
\epsfbox[18 144 592 718] {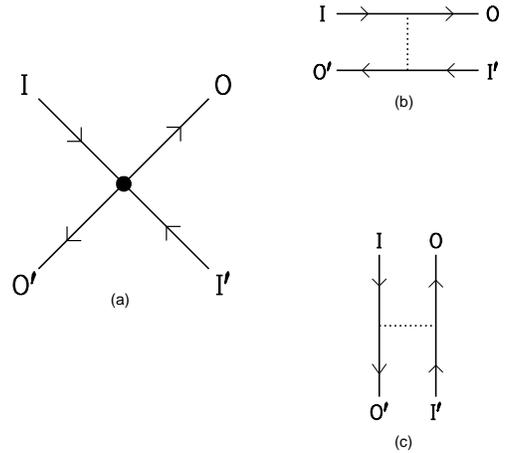}
\caption[]
{\label{vertex} (a) The $S$-matrix relates incoming to outgoing flux
amplitudes (see eq. (\ref{Smat})).  This information may alternately be
expressed as a transfer matrix $M$ relating left and right flux amplitudes
(b), or $N$ relating top and bottom flux amplitudes (c).}
\end{figure}
This $S$-matrix may be recast in the form of a ``left-to-right''
transfer matrix $M$, which relates flux amplitudes
$\cO$ and $\cI'$ to $\cI$ and $\cO'$:
\begin{eqnarray*}
\pmatrix{\cO\cr\cI'}&=&M\pmatrix{\cI\cr\cO'}\\
M&=&\pmatrix{\sec\theta &\tan\theta\cr\tan\theta &\sec\theta\cr}\ .
\end{eqnarray*}
The transmission coefficient $T$ relating the incident
flux $|\cI|^2$ to the outgoing flux $|\cO|^2$ is $T=\cos^2\theta$,
and the reflection coefficient is $R=1-T=\sin^2\theta$.

Consider now the ``series'' transmission through two consecutive
scatterers.  Multiplying transfer matrices, and taking into account
the random phases accrued in between scatterers, one obtains a composite
transfer matrix
\begin{displaymath}
{\wt M}=\pmatrix{\sec\theta_2 &\tan\theta_2\cr
\tan\theta_2 &\sec\theta_2\cr}
\pmatrix{e^{i\alpha}&0\cr 0&e^{i\beta}\cr}
\pmatrix{\sec\theta_1 &\tan\theta_1\cr\tan\theta_1 &\sec\theta_1\cr}
\end{displaymath}
and from $\sec^2\theta'=|{\wt M}_{1,1}|^2=1/T'$, a transmission coefficient
\begin{equation}
{1\over T'}={1\over T_1}\cdot{1\over T_2}\cdot\left[1+2\sqrt{R_1 R_2}\cos\phi
 + R_1 R_2\right]\ ,
\label{EQAnd80}\end{equation}
where $\phi=\alpha-\beta$ is random.  Precisely this calculation was
done by Anderson \etal\ \cite{And80} in their study of one-dimensional
localization (see also refs.~\cite{Mel86,Sha87}.  Averaging $\ln T$ over
the angle $\phi$, one obtains
\begin{equation}
\langle\ln T'\rangle = b\,\langle\ln T\rangle
\label{seriesT}
\end{equation}
for $b$ scatterers in series.  Thus, $\ln T$ is driven to increasingly
negative values -- this is the essence of one-dimensional localization.
\begin{figure} [t]
\centering
\leavevmode
\epsfxsize=8cm
\epsfysize=8cm
\epsfbox[18 144 592 718] {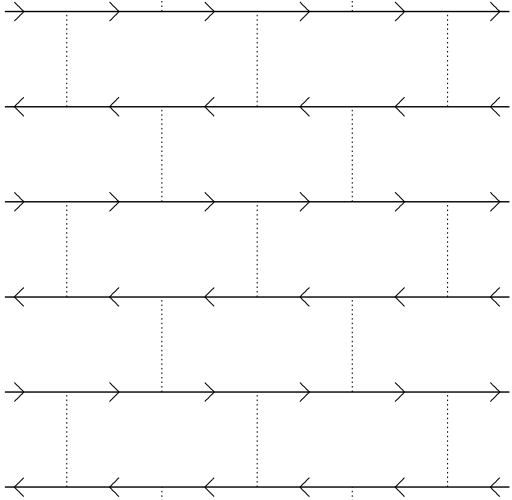}
\caption[]
{\label{bhor} The horizontal `bricklayer's representation' of the network
model. Quantum mechanical particles propagate in the directions of the arrows;
scattering processes are represented by the dotted lines.  The corresponding
vertical representation is rotated by $90^\circ$.}
\end{figure}

Equivalently, though, one may view the propagation as ``top-to-bottom'', and
define a transfer matrix $N$ relating $\cO'$ and $\cI'$
to $\cI$ and $\cO$:
\begin{eqnarray*}
\pmatrix{\cO'\cr\cI'}&=&N\pmatrix{\cI\cr\cO}\\
N&=&\pmatrix{-\csc\theta &\ctn\theta\cr-\ctn\theta &\csc\theta\cr}\ .
\end{eqnarray*}
In this case, the roles of $T$ and $R$ are reversed: $T=\sin^2\theta$,
$R=\cos^2\theta$, and two scatterers in parallel give
\begin{equation}
{1\over R'}={1\over R_1}\cdot{1\over R_2}\cdot\left[1+2\sqrt{T_1 T_2}\cos\phi
 + T_1 T_2\right]\ ,
\end{equation}
with
\begin{displaymath}
\langle\ln R'\rangle = b\,\langle\ln R\rangle
\end{displaymath}
for $b$ parallel scatters.  In this case the reflection amplitude is driven
to zero.  Now in the network model, both series as well as parallel propagation
occur, and in our viewpoint it is a competition between these two
one-dimensional localization mechanisms which leads to a critical
point corresponding to the quantum Hall transition.  It is, however,
impossible to neatly separate the two modes of propagation,
and we shall have to resort to approximation schemes, such as the
Migdal-Kadanoff approach, or to a modification of the original model,
such as hierarchical network constructions, in order to make progress.
\begin{figure} [t]
\centering
\leavevmode
\epsfxsize=8cm
\epsfysize=8cm
\epsfbox[18 144 592 718] {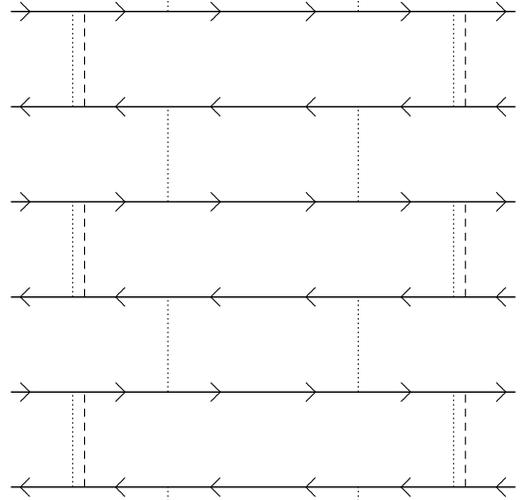}
\caption[]
{\label{bhor2} The first step in the $b=2$ Migdal-Kadanoff renormalization of
the network model.  Compare with fig.~\ref{bhor}.  Dashed lines indicate
bonds which have been shifted to the left.  Dotted lines indicate unshifted
bonds.}
\end{figure}

\section{Migdal-Kadanoff Approach}\label{secMK}

\begin{figure} [t]
\centering
\leavevmode
\epsfxsize=9cm
\epsfysize=9cm
\epsfbox[18 144 592 718] {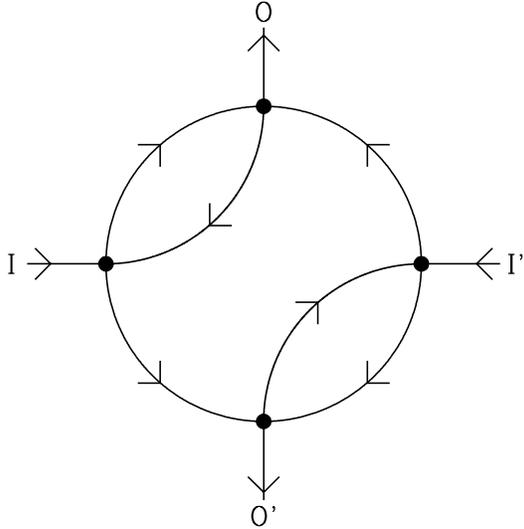}
\caption[]
{\label{SP} The composite vertex in the $b=2$ Migdal-Kadanoff renormalization,
which corresponds to series followed by parallel propagation.}
\end{figure}

\begin{figure} [t]
\centering
\leavevmode
\epsfxsize=9cm
\epsfysize=9cm
\epsfbox[18 144 592 718] {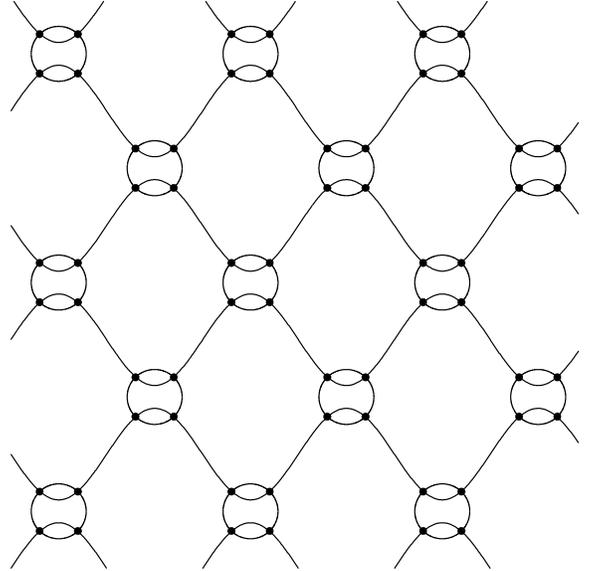}
\caption[]
{\label{MK2x2} After shifting in both the horizontal and vertical directions,
each vertex in the network represents a composite of $b^2$ vertices.  The case
$b=2$ is shown here.}
\end{figure}

In order to apply the Migdal-Kadanoff bond shifting scheme to the
network model, we begin by graphically representing the elementary scattering
process as a kind of interaction vertex, as depicted in fig.~\ref{vertex}.
Thus, the entire network may be represented by an infinite sequence of 
parallel lines on which currents flow in alternating directions,
and backscattering processes which divert flux from one line to one
of its neighbors (see fig.~\ref{bhor}).  We call this the `bricklayer's
representation' of the network model, for obvious reasons, and further
describe the propagation as being either `horizontal' or `vertical'.
To effect a rescaling by a factor $b$, we start with the (horizontal)
bricklayer's representation, shown in fig.~\ref{bhor}.
In every other row, we then shift $b-1$ out of every $b$ interaction lines
so as to form a new network in which each renormalized vertex represents $b$
bare vertices in series.  This is depicted for $b=2$ in fig.~\ref{bhor2}.
We then rescale horizontal distances by $b$.
Were we to proceed in this manner, we would obtain the one-dimensional
localization of ref.~\cite{And80}.  Instead, we view the network again
in the bricklayer's representation, this time with vertical propagation,
and perform a second bond-shifting operation.  This replaces each
scattering vertex with an effective composite vertex, shown in fig.~\ref{SP}.
The new network resembles that of fig.~\ref{MK2x2}.

To see how this leads to critical behavior, consider the behavior of
the {\it typical} transmission coefficient, $\exp\langle\ln T\rangle$, under
such a transformation.  From eq. (\ref{seriesT}), we obtain
\begin{equation}
T'=1-(1-T^b)^b\equiv f(T;b)\ ,
\label{Tprime}
\end{equation}
where we implicitly are working with typical values.  For the case $b=2$,
we have
\begin{displaymath}
f(T;b=2)=2T^2-T^4\ ,
\end{displaymath}
which has two stable fixed points at $T^*=0$ and $T^*=1$, and one
unstable fixed point at $T^*=\half(\sqrt{5}-1)\approx 0.618$.
Linearizing about the unstable fixed point, we obtain an eigenvalue
$(\pz f(T;b=2)/\pz T)\nd_{T^*}=6-2\sqrt{5}$, corresponding to a
localization length exponent of $\nu=\ln b/\ln\lambda\approx 1.635$.
Note that $T^*\neq\half$, a consequence of the order in which the bond
shifting was performed: the composite vertex corresponds to series
followed by parallel propagation (note the composite vertex of
fig.~\ref{SP} are not symmetric under $90^\circ$ rotations).  We could
equally well have chosen parallel followed by series propagation,
which would switch the roles of $T$ and $R$.  We note that the RG
equation (\ref{Tprime}) and its counterpart with $T$ replaced by $1-T$
coincide with two RG equations obtained in a Migdal-Kadanoff approach
to classical bond percolation \cite{Kir77}.  There, the bond occupation
probability $p$ plays the role of $T$, and the MK bond shifting which
leads to the composition of quantum resistors in the network model
corresponds to multiplication of bond occupation probabilities.

\begin{figure} [t]
\centering
\leavevmode
\epsfxsize=9cm
\epsfysize=9cm
\epsfbox[18 144 592 718] {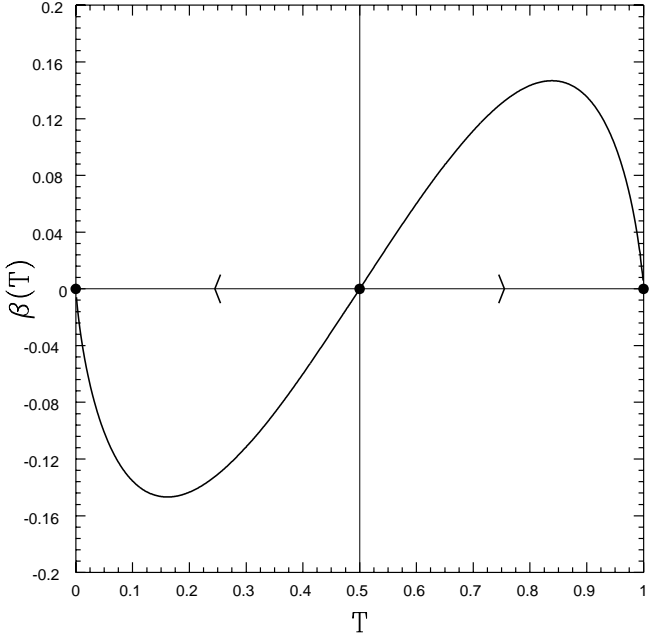}
\caption[]
{\label{MKbeta} The $\beta$-function resulting from the infinitesimal
Migdal-Kadanoff renormalization.  $T=0$ and $T=1$ are attractive fixed points;
the critical point lies at $T=\half$.  Arrowheads point in the direction of
RG flow.}
\end{figure}

If we set $b=1+\zeta$, where $\zeta\to 0$, we obtain equations for an
`infinitesimal' Migdal-Kadanoff renormalization:
\begin{equation}
T'=T+\zeta\left[T\ln T-(1-T)\ln (1-T)\right]+\cO(\zeta^2)\ .
\label{infMK}
\end{equation}
The infinitesimal MK transformation has a fixed point at $T^*=\half$,
and an eigenvalue $\lambda=2(1-\ln 2)\zeta$, corresponding to a critical
exponent of $\nu=1/2(1-\ln 2)\approx 1.629$.  The MK beta function is then
\begin{eqnarray}
\beta(T)&=&{d\over d\zeta}\bigg|_{\zeta=0} f(T;1+\zeta)\nonumber\\
&=&T\ln T - (1-T)\ln (1-T)\ ;
\label{MKbf}
\end{eqnarray}
this is shown in fig.~\ref{MKbeta}.

In statistical mechanics applications, such as the Ising model, one can
define a new Hamiltonian $\cH'$ in which bonds have been shifted according
to the MK prescription.  When computing the partition function, one can
in principle perform thermodynamic perturbation theory in $(\cH'-\cH)$ in
order to systematically improve upon the MK procedure.  We know of no such
systematic improvement for our scheme, nor do we have any sort of
reliable estimate for the errors involved in the calculation of $\nu$.

\section{Hierarchical Lattices}\label{secHL}

\begin{figure} [t]
\centering
\leavevmode
\epsfxsize=9cm
\epsfysize=9cm
\epsfbox[18 144 592 718] {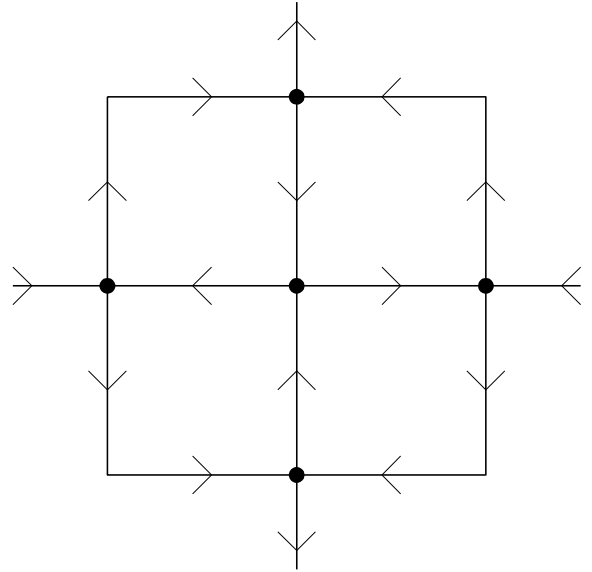}
\caption[]
{\label{HLcell} An $M=3$, $V=5$ elementary cell on which a hierarchical
lattice is based.}
\end{figure}

\begin{figure} [t]
\centering
\leavevmode
\epsfxsize=9cm
\epsfysize=9cm
\epsfbox[18 144 592 718] {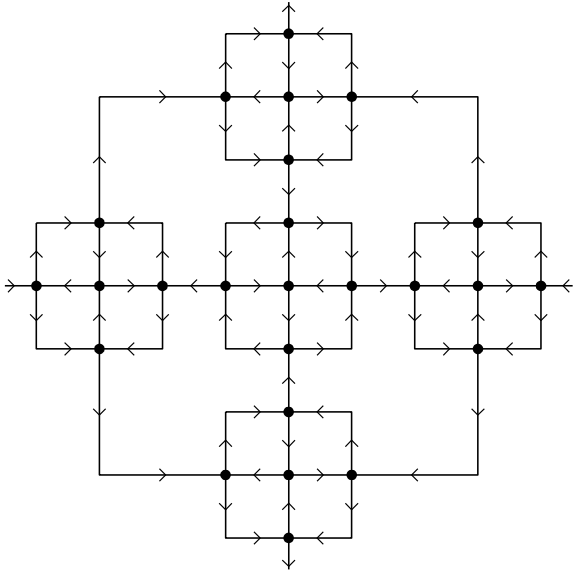}
\caption[]
{\label{HL} The hierarchical lattice at the second level of construction,
beginning with the $M=3$, $V=5$ cell of fig.~\ref{HLcell}.  There are
$V^2=25$ vertices in this $M^2\times M^2=9\times 9$ supercell.}
\end{figure}

A related approach to the problem involves the construction of hierarchical
lattices.  Consider, for example, the scattering unit of fig.~\ref{HLcell},
which contains $V=5$ vertices in an area $M^2$; in units of the distance
between vertices, we have $M=3$.  Now replace each of the vertices with
a replica of the original cell, forming the structure shown in fig.~\ref{HL}.
Repeating the process $n$ times generates a hierarchical structure with
$V^n$ vertices contained in a square of side length $M^n$.  The Hausdorff
dimension of this hierarchical lattice is $\DH=\ln V/\ln M\approx 1.465$.
From $V(M)=\half(M^2+1)$, one has $\lim_{M\to\infty}\DH(M)=2$.
The renormalization flow of the transmission coefficient and the multifractal
spectrum $f(\alpha)$ was calculated in ref.~\cite{Wey97} for the cases
$M=3$, $M=5$, and $M=9$; the results for $\nu$ and $f(\alpha)$ rapidly
converge to the network model results even for such modest sizes.
The RG equations for the case $M=3$ are identical to those obtained in
ref.~\cite{Gal97}, where a different interpretation is adopted.

If one replaces the central scatterer in fig.~\ref{HLcell} with one for
which $T=1$, one recovers the four site scattering unit of fig.~\ref{SP}.
One can use this as the fundamental unit of a hierarchical construction,
and the results differ from those of the previous section only in that
the linear dimension is taken to be $M=2b-1$ rather than $b$.  The Hausdorff
dimension of the hierarchical lattice is $\DH=\ln b^2/\ln M$ ($=\ln 4/\ln 3$ 
for $b=2$), whereas the lattice of fig.~\ref{MK2x2} is fully two-dimensional.
The exponent $\nu$ now is changed:
$\nu=\ln M/\ln\lambda=\ln 3/\ln(6-2\sqrt{5})\approx 2.592$.
Generalizations to other values of $b$ are straightforward:
iterating eq. (\ref{Tprime}) $n$ times, we obtain the flow equation
\begin{equation}
T(M^n\ell_0)=f^{[n]}(T(\ell_0);b=\half M + \half)\ ,
\label{RGflow}
\end{equation}
where $f^{[n]}$ is the iterated function satisfying
\begin{displaymath}
f^{[n+1]}(T;b)=f\left(f^{[n]}(T;b)\right)
\end{displaymath}
and $f^{[1]}\equiv f$.  Since $f(T;b)$ is monotonically increasing from
$f(0;b)=0$ to $f(1;b)=1$, there are always three fixed points on the interval
$T\in [0,1]$: $T=0$ and $T=1$ (both stable), and a nontrivial unstable
fixed point with $0 < T^* <1$.  The critical exponent $\nu$ is
\begin{equation}
\nu={\ln M\over\ln f'(T^*;\half M+\half)}
\end{equation}
where $f'(T;b)=df(T;b)/dT$.

\begin{figure} [t]
\centering
\leavevmode
\epsfxsize=9cm
\epsfysize=9cm
\epsfbox[18 144 592 718] {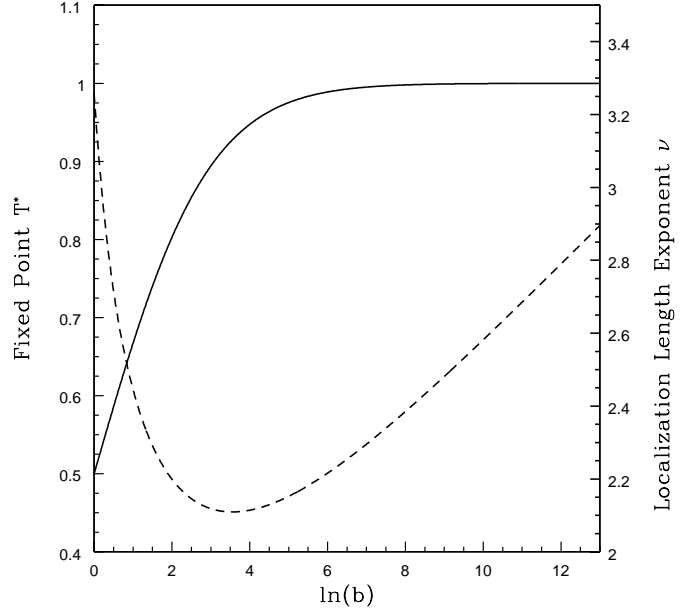}
\caption[]
{\label{NU} Results of the hierarchical lattice renormalization of
eq. (\ref{RGflow}) for the fixed point transmission coefficient, $T^*$ (solid
curve), and the localization length exponent, $\nu$ (dashed curve),
{\it versus\/} $\ln b$.  The linear dimension is $M=2b-1$.}
\end{figure}

Whereas the Hausdorff dimension $\DH$ and fixed point both increase
monotonically, with
\begin{eqnarray*}
\DH(b=1)=1\quad &,&\quad \DH(b=\infty)=2\\
T^*(b=1)=\half\quad &,&\quad T^*(b=\infty)=1\ ,
\end{eqnarray*}
the critical exponent $\nu$ exhibits a minimum for $b\approx 35$,
where $\nu\approx 2.11$.  We find that $\nu$ diverges weakly (as
$\ln b/\ln\ln b$) in the $b\to\infty$ limit.  These results are plotted
in fig.~\ref{NU}.

\section{Renormalization Group for Distributions}\label{secrgd}

In section \ref{secMK}, we derived RG flows for the typical transmission
coefficient, defined by $T_{\rm typ}\equiv \exp\langle\ln T\rangle$.
Now we will concern ourselves with the RG flow for the entire distribution
$P(T)$.  We will choose the simplest nontrivial unit cell, namely that of
fig.~\ref{SP}, in which two composite scatterers, each of which is two
bare scatterers in series, are placed in parallel.  

Starting with a distribution function $P(T)$, we derive the intermediate
distribution $P\nd_{\rm int}$ after taking two scatterers in series:
\begin{eqnarray}
P\nd_{\rm int}(T)&=&\int_0^1\!\!\!dT_1\,P(T_1)\int_0^1\!\!\!dT_2\,P(T_2)\times
\nonumber\\
&&\quad\times\int_0^\pi\!\!{d\phi\over\pi}\,\delta(T-f(T_1,T_2,\cos\phi))
\label{EQDIS1}\end{eqnarray}
with
\begin{displaymath}
f(T_1,T_2,\mu)={T_1\,T_2\over 1 + (1-T_1)(1-T_2)+2\mu
\sqrt{(1-T_1)(1-T_2)}}\ .
\end{displaymath}
We now combine two of these composite units
in parallel, obtaining the renormalized transmission coefficient distribution
\begin{eqnarray}
{\wt P}(T)&=&\int_0^1\!\!\!dR_1\,P\nd_{\rm int}(1-R_1)\int_0^1\!\!\!dR_2\,
P\nd_{\rm int}(1-R_2)\times\nonumber\\
&&\quad\times\int_0^\pi\!\!{d\phi\over\pi}\,\delta(1-T-f(R_1,R_2,\cos\phi))\ .
\label{EQDIS2}\end{eqnarray}
This is equivalent to the following:
\begin{eqnarray*}
P\nd_{\rm int}(T)&=&{1\over 2\pi T^2}\int_0^1\!\!\!dT_1\int_0^1\!\!\!dT_2\,
(T_1)\,P(T_2)\,{\Theta(1-|A|)\over\sqrt{1-A^2}}\times\\
&&\qquad\times{T_1\,T_2\over\sqrt{(1-T_1)(1-T_2)}}\\
&&\vphantom{\Sigma}\\
A&=&{T_1\,T_2-T(1+(1-T_1)(1-T_2))\over 2T\sqrt{(1-T_1)(1-T_2)}}\\
\end{eqnarray*}
followed by
\vfill
\begin{eqnarray*}
{\wt P}(T)&=&{1\over 2\pi (1-T)^2}\int_0^1\!\!\!dT_1\int_0^1\!\!\!dT_2\,
P\nd_{\rm int}(T_1)\,P\nd_{\rm int}(T_2)\times\\
&&\quad\times{\Theta(1-|B|)\over\sqrt{1-B^2}}
\,{(1-T_1)(1-T_2)\over\sqrt{T_1\,T_2}}\\
&&\vphantom{\Sigma}\\
B&=&{(1-T_1)(1-T_2)-(1-T)(1+T_1\,T_2)\over 2(1-T)\sqrt{T_1\,T_2}}\ .
\end{eqnarray*}
Note that $B(T_1,T_2;T)=A(1-T_1,1-T_2;1-T)$.

\begin{figure} [t]
\centering
\leavevmode
\epsfxsize=9cm
\epsfysize=9cm
\epsfbox[18 144 592 718] {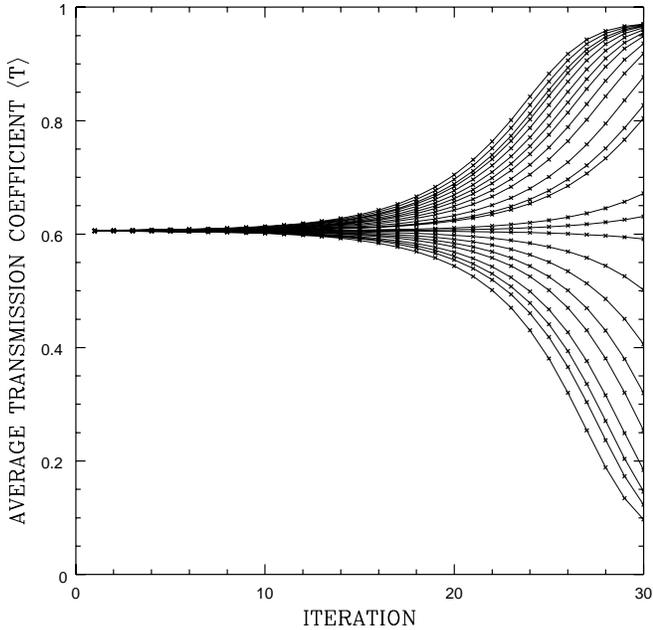}
\caption[]
{\label{DFLOW} The RG flow of the distribution $P(T)$.  The average transmission
coefficient $\langle T\rangle$ is plotted {\it versus\/} iteration for various
initial conditions.  The initial value of $\Delta T$ was held fixed as
$\langle T\rangle\nd_{\rm init}$ was varied between $0.605$ and $0.606$.  The
critical point was found to lie at $\langle T\rangle\nd_{\rm init}\approx
0.60536$.}
\end{figure}

\begin{figure} [t]
\centering
\leavevmode
\epsfxsize=9cm
\epsfysize=9cm
\epsfbox[18 144 592 718] {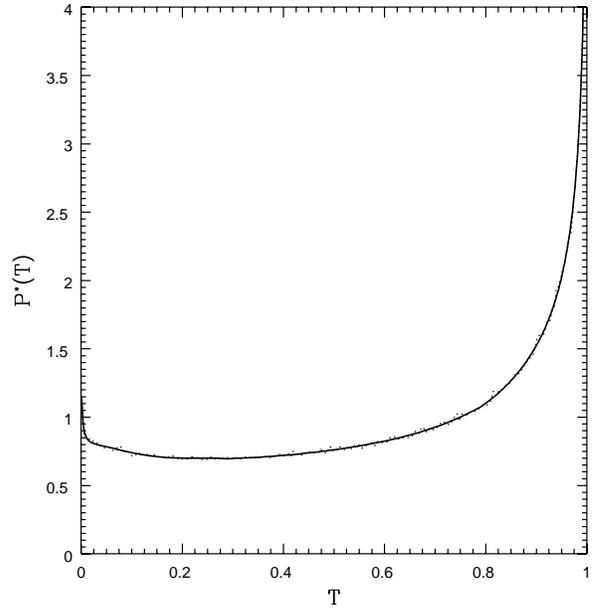}
\caption[]
{\label{FD} The fixed distribution $P^*(T)$ obtained by iteration of the
RG flow equations.  Shown are results of a numerical integration (dotted
curve) and a smooth fit to the numerical data (solid curve).}
\end{figure}

We have numerically iterated the distribution according to these
equations.  Since we expect a single relevant scaling variable
controlling the RG flow of $P(T)$, we parameterize the initial
distribution by its average $\langle T\rangle\nd_{\rm init}$ and a
width $(\Delta T)\nd_{\rm init}= \sqrt{\langle T^2-\langle T\rangle^2
\rangle}$.  By varying $\langle T\rangle\nd_{\rm init}$ alone, we
drive the system through its critical point.  This is shown in
fig.~\ref{DFLOW}, where the iterated average transmission coefficient
is plotted for several initial values of $\langle T\rangle$.  (We used
an initial distribution which was a sum of linear plus exponentially
increasing terms, though the precise details are presumably
unimportant.  Keeping the initial width fixed and varying $\langle
T\rangle\nd_{\rm init}$ between $0.605$ and $0.606$, we found the
critical point $\langle T\rangle^*$ at $\langle T\rangle\nd_{\rm
init} \approx 0.60536$.)  The fixed point distribution itself is
plotted in fig.~\ref{FD}.

\begin{figure}[t]
\centering
\leavevmode
\epsfxsize=7cm
\epsfysize=7cm
\epsfbox[0 0 450  400] {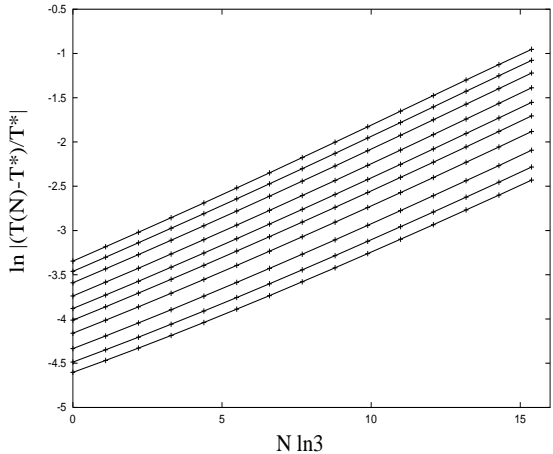}
\caption[]
{\label{NUDIS} To derive $\nu$ from eq.~\ref{NUeqn}, we plot the logarithm
of the left hand side {\it versus\/} $N\ln M$.  The slope is then $\nu^{-1}$.
Here we show results for $M=3$ discussed in sec.~\ref{secrgd}; we find
$\nu\approx 6.6$, in poor agreement with the accepted value $\nu\approx
\frac{7}{3}$.}
\end{figure}

Within the hierarchical lattice approach, each iteration rescales the
linear system size by a factor $M=3$.  Analyzing the iterated average
transmission coefficient $\langle T\rangle$ as a function of the
number $N$ of iterations allows for the computation of $\nu$ from
\begin{equation}
\langle T\rangle\nd_N -\langle T\rangle^* \propto M^{N/\nu}\ .
\label{NUeqn}
\end{equation} 
In fig.~\ref{NUDIS} we plot $\ln |(\langle T\rangle_N -\langle
T\rangle^*)/\langle T\rangle^*|$ {\it versus\/} $N\ln M$; we find
$\nu\approx 6.6$.  Note that within the MK scheme, the rescaling
factor is $b=2$, consequently $\nu\approx 4.2$.

The iteration of the full distribution function $P(T)$ according to
eqs.~(\ref{EQDIS1},\ref{EQDIS2}) yields an exact RG flow for $\PL$ 
for the  four site scattering unit of fig.~\ref{SP}.  However,
the iteration could be performed only numerically.  We conclude this
section with a brief account of an analytical method to obtain $P(T)$.
In the context of one-dimensional localization eq.~(\ref{EQAnd80}) has
been used to derive a Fokker-Planck equation ({\it cf.\/}~\cite{Ris89})
for the distribution function $\PL$.  One identifies $T_1\equiv T(L)$ with the
transmission coefficient of a one-dimensional resistor of length $L$ placed
in series with a resistor of infinitesimal length and transmission
coefficient $T_2\equiv T(\delta L)$.  In a regime of `local weak scattering',
one can identify the elastic mean free path $l_\rme$ by
$l_{\rm e}=\delta L/\langle R(\delta L)\rangle$ and derive the following
Fokker-Planck equation by standard methods ({\it cf.\/}~\cite{Sha87} and
references therein):
\begin{equation}
l_\rme\,{\pz\PL\over\pz L}={\pz\over\pz T}\left\{
(1-T)\,{\pz\over\pz T}\,\left( T^2\PL\right)\right\}
\label{Fok2}   
\end{equation}
This is an equation describing one-dimensional localization.  It turns out
that $l_{\rme}$ sets the scale for the localization length,
\ie\ $\langle -\ln T\rangle\nd_L = L/l_\rme$.

\begin{figure}[t]
\centering
\leavevmode
\epsfxsize=7cm
\epsfysize=7cm
\epsfbox[0 0 350 400] {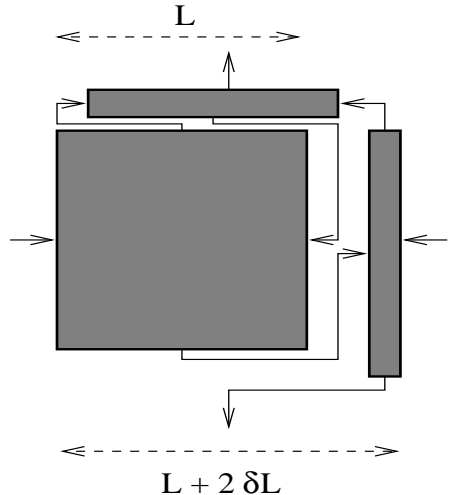}
\caption[]
{\label{FOKKER} Construction of a network by adding infinitesimal networks to
a square network of linear size $L$. The infinitesimal networks are
characterized by a small extension $\delta L$ in  one direction.
The network is composed
according to series and parallel composition laws.}
\end{figure}

We now try to extend this one-dimensional scheme to the network model.
Our two-dimensional iteration scheme is graphically depicted in
fig.~\ref{FOKKER}.  We begin with an $L\times L$ array and attach to it
two infinitesimal rectangles of size $L\times\delta L$
and $\delta L\times (L+\delta L)$, respectively.  Transport then takes
place according to the usual series and parallel composition laws.
The main obstacle in deriving a Fokker-Planck equation for this model
is due to the fact that we do not know {\it a priori\/} how the transmission
coefficients of the infinitesimal blocks behave. To make progress, we
proceed in analogy to the one-dimensional case and {\it assume}
\begin{eqnarray}
\langle R(L,\delta L)\rangle &=& \langle -\ln T \rangle\nd_L\cdot
(\delta\! L/L)\nonumber\\ 
\langle T(\delta L, L+\delta L)\rangle &=& \langle -\ln (1-T) \rangle\nd_L
\cdot(\delta\! L/L)\ .
\label{FOK4} 
\end{eqnarray}
With this assumption the resulting Fokker-Planck equation reads
\begin{equation}
{\pz\over\pz L}\PL=-{\pz\over\pz T}{\cal J}(T;L)\ ,
\label{Fok5}
\end{equation}
where the `current' is given by
\begin{eqnarray}
{\cal J}(T;L)&=&\langle \ln T\rangle\nd_L\, {1-T\over 2L}\,{\pz\over\pz T}
\left\{T^2\PL\right\}]+\label{Fok6}\\
&&\quad+\,\langle\ln(1-T)\rangle\nd_L\, {T\over 2L}\, {\pz\over\pz T}
\left\{(1-T)^2\PL\right\}.\nonumber
\end{eqnarray}
Without going into details we report on the results that one can
obtain from the Fokker-Planck equation (\ref{Fok5}). 
We find that there are three fixed point distributions: $\delta (T)$,
$\delta (1-T)$, and a uniform critical distribution $P^*(T) \equiv 1$.
To extract the critical exponent $\nu$  of the localization length
we took  $X\nd_L=  \langle \ln (1-T)\rangle\nd_L - \langle \ln T\rangle\nd_L$
as a scaling variable which has a fixed point value $X^*=0$.  Linearizing the
flow equation around the fixed point distribution gives rise to
a beta function for $X\nd_L$ which reads
\begin{equation}
{d X\nd_L\over d \ln L}= X\nd_L\ .
\label{Fok8}
\end{equation}
Consequently, the critical exponent is $\nu =1$. Although the
Fokker-Planck approach to the network model is able to describe the
correct qualitative physics of the localization-delocalization
transition, it suffers from an ambiguity in modeling the average transmission
coefficients of the infinitesimal blocks.  Adopting a different
model can change the critical exponent $\nu$. In addition, the
composition laws on which the Fokker-Planck approach is based 
do not lead to the full Chalker-Coddington network.

\section{Conclusion}

In this paper we developed a simple approach to the quantum Hall
transition.  We considered a number of hierarchical network models related
to the Chalker-Coddington model of quantum percolation.  The basic ingredients
of our networks are series and parallel compositions of quantum resistors;
their hierarchical nature allows us to analyze their critical properties
using simple real space renormalization techniques.
In particular we studied network models that can be 
interpreted as resulting either from Migdal-Kadanoff bond-shifting
scheme or from the construction of hierarchical lattices of Hausdorff
dimension $\DH<2$.  The different models are labeled by an integer
$b$ where $b^2$ is the number of resistors that form an elementary cell.

In sec.~\ref{secrgd} we calculated the flow of the distribution function
$\PL$ for the transmission coefficient $T$ (`two-point-conductance')
in the case $b=2$.  This was done by means of numerically iterating the
renormalization group equations.
We found the flow to have three fixed points, two of which correspond
to localization ($T\equiv 0$, $T\equiv 1$).  The third fixed distribution
$P^*(T)$ corresponds to the quantum critical point, where delocalization
occurs.  The critical point distribution $P^*(T)$ turned out to be
very broad and the flow of $\PL$ in the vicinity of $P^*(T)$ is governed
by one-parameter scaling.  The critical exponent $\nu$ was also determined
and found in poor agreement with accepted values.

As a further simplification we derived renormalization schemes for the typical
transmission coefficient, $T_{\rm typ}\equiv\exp \langle \ln T\rangle$.
Here, we simply averaged over the random link phases, neglecting variations
in the individual transmission coefficients.  We obtained closed RG equations
for $T_{\rm typ}$ (secs.~\ref{secMK},\ref{secHL}).
From these equations we derived fixed point values for 
$T_{\rm typ}^*(b)$ and critical localization length exponents $\nu\nd_b$
for arbitrary $b>1$.  For small $b$ the notion of typical transmission
coefficient as a substitute for a whole distribution (which is broad)
is dubious and it is no surprise that $\nu\nd_{b=2}$ differs from the value
obtained by iterating the whole distribution function.
However, the notion of typical transmission coefficient
is more reliable for large $b$ where
the composition of $b$ resistors in series favors the formation of
log-normal type distributions.
On the other hand, for large $b$ the resulting hierarchical network has
little in common with the original Chalker-Coddington network.  Nevertheless,
 the slow variation of $\nu\nd_b$ with $b$ shows that the results for
intermediate $b$ are not too far from that of the original network model.

All models that we considered exhibit a 
localization-delocalization transition which results from a
competition between two one-dimensional quantum mechanical
localization mechanisms.  The models result from uncontrolled approximations
to the network model, but have at least some pedagogical value.
In contrast to the well-known classical percolation model for quantum
Hall transitions, our models are essentially quantum mechanical.
Furthermore, the concept of real space renormalization can be extended
to improve quantitative results (see refs. ~\cite{Gal97},\cite{Wey97}).

While this manuscript was in the final stage of preparation, we
discovered the work of ref.~\cite{Gal97} in which a similar
real space renormalization group approach to the Chalker-Coddington
model is developed.  In that work an elementary $M=3$, $V=5$ cell
(see fig.~\ref{HLcell}) replaces each scatterer in the renormalization
step.  (In addition, the rescaling factor in \cite{Gal97} is $b=2$ rather
than $M=3$.)  In our work the central scatterer of this cell is replaced
with one for which $T=1$ (or $R=1$), resulting in the simple series-parallel
composition laws discussed above.

\section{Acknowledgements}\label{secACK}

DPA and MJ would like to thank the Physics Department at Technion,
where this work was started.  DPA also thanks the Lady Davis Fellowship
Trust and the National Science Foundation, grant NSF DMR-91-13631, for
partial support.  MJ also thanks the \hbox{MINERVA} foundation and the
Sonderforschungsbereich 341 of the Deutsche Forschungsgemeinschaft
for partial support.
We are grateful to A. Weymer for discussions
and for assistance in computing the exponent in sec.~\ref{secrgd}.


\end{document}